\documentclass{webofc}
\usepackage[varg]{txfonts}
\usepackage{hyperref}

\usepackage{lineno}
\usepackage{cleveref}

\crefformat{section}{\S#2#1#3} 
\crefformat{subsection}{\S#2#1#3}
\crefformat{subsubsection}{\S#2#1#3}

\usepackage[normalem]{ulem}

\newcommand{\textpack}[1]{\textsc{#1}}

\usepackage{xcolor}
\selectcolormodel{rgb}
\definecolor{LHCb dark}{rgb}{0.0000,0.3412,0.6549}
\definecolor{UC red}{rgb}{0.8196,0.1176,0.2314} 
\definecolor{brickred}{rgb}{0.8, 0.25, 0.33}
\begin{document}
\title{Progress in developing a hybrid deep learning algorithm for identifying and locating primary vertices}
%
%

\author{\firstname{Simon} \lastname{Akar}\inst{1}\fnsep
 \thanks{\email{simon.akar@cern.ch}}
    \and
        \firstname{Gowtham} \lastname{Atluri}\inst{1}
     \and
        \firstname{Thomas} \lastname{Boettcher}\inst{1}
     \and
        \firstname{Michael} \lastname{Peters}\inst{1}
     \and
        \firstname{Henry} \lastname{Schreiner}\inst{2}
     \and
        \firstname{Michael} \lastname{Sokoloff}\inst{1}\fnsep
        \thanks{\email{mike.sokoloff@uc.edu}}
     \and
        \firstname{Marian} \lastname{Stahl}\inst{1}
      \and
        \firstname{William} \lastname{Tepe}\inst{1}
      \and
        \firstname{Constantin} \lastname{W}eisser\inst{3}
      \and
        \firstname{Mike} \lastname{Williams}\inst{3}
}

\institute{University of Cincinnati 
\and
 Princeton University
\and
  Massachusetts Institute of Technology}

\abstract{
The locations of proton-proton collision points in LHC experiments are called primary vertices (PVs). 
Preliminary results of a hybrid deep learning  algorithm for identifying and locating these, targeting the Run 3 incarnation of LHCb, have been described at conferences in 2019 and 2020. 
In the past year we have made significant progress in a variety of related areas. 
Using two newer Kernel Density Estimators (KDEs) as input feature sets improves the fidelity of the models, as does using full LHCb simulation rather than the ``toy Monte Carlo'' originally (and still) used to develop models. 
We have also built a deep learning model to calculate the KDEs from track information. 
Connecting a {\tt tracks-to-KDE} model to a {\tt KDE-to-hists} model used to find PVs provides a  proof-of-concept that a single deep learning model can use track information to find PVs with high efficiency and high fidelity. 
We have studied a variety of models systematically to understand how variations in their architectures  affect performance. 
While the studies reported here are specific to the LHCb geometry and operating conditions, the results suggest that the same approach could be used by the ATLAS and CMS experiments.}
\maketitle
%
\section{Introduction}
\label{intro}
The LHCb experiment is currently being upgraded for the planned start of Run 3 of the LHC in 2022. 
It will record proton-proton collision data at five times the instantaneous luminosity of Run 2. 
The average number of visible primary vertices (PVs), proton-proton collision in the detector closest to the beam-crossing region, will increase from 1.1 to 5.6. 
Building on the success of Run 2, the experiment will move to a pure software data ingestion and trigger system, eliminating the Level 0 hardware trigger altogether~\cite{LHCbCollaboration:2014vzo}. 
A conventional PV finding algorithm~\cite{Reiss:2749592, Reiss:CtD2020} that satisfies all requirements defined in the Trigger Technical Design Report~\cite{Aaij:2018jht} serves as the baseline. 
In parallel, we have been developing a hybrid machine learning algorithm, designed to run in the initial stage of the LHCb upgrade trigger.

\begin{figure}
\centering
\includegraphics[width=0.65\textwidth,clip]{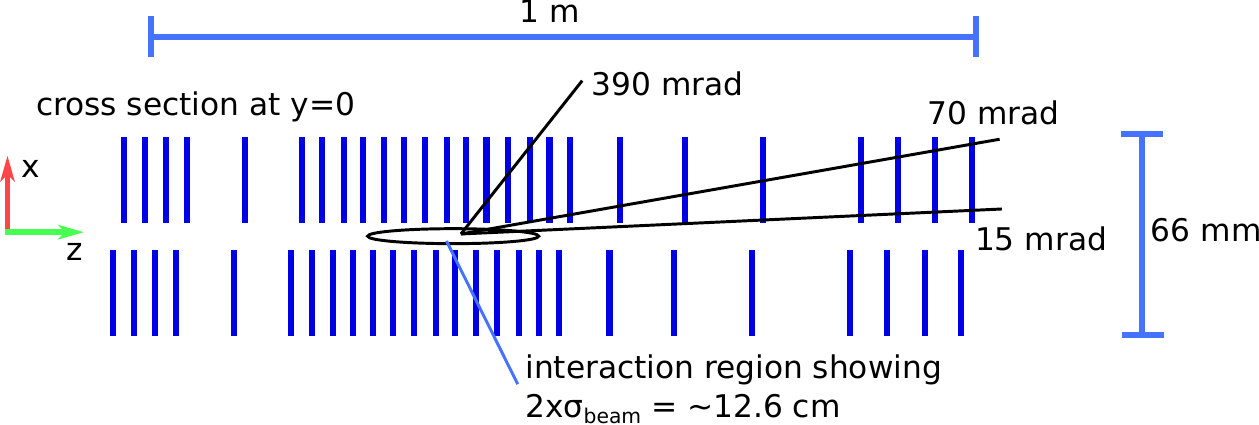}
\caption{This diagram illustrates the luminous region of the LHCb experiment and the Vertex Locator (VELO), the high precision silicon pixel detector that surrounds it.}
\label{fig:velo}
\end{figure}

A cartoon illustrating the Run 3 Vertex Locator (VELO)~\cite{LHCbCollaboration:2013bkh} is shown in Fig.~\ref{fig:velo}. 
Approximately 41 million pixels, $ 55 \times 55 \,\mu {\rm m}^2 $ each, will populate 26 circular discs oriented perpendicular to the beamline. 
The LHCb detector is a forward spectrometer with the remaining elements to the right of the VELO in this view. 
In the longitudinal direction, the luminous region can be characterized by the ellipse labeled as the interaction region. 
In the transverse directions, most PVs are produced within 40 $ \mu $m of the center of the beamline. 
An error ellipse with its minor axis drawn to the vertical scale of the figure would be a factor of 100 thinner than that shown. 
The typical longitudinal resolution of a PV reconstructed following full tracking and fitting is $ 40 - 200 \, \mu $m (depending primarily on track multiplicity), and the typical transverse resolution is  $ \sim 10 \, \mu $m.
They also allow tracking and PV-finding algorithms to use these constraints to execute quickly and efficiently. 
Of interest for the work presented here, PV-finding algorithms can use track parameters evaluated at their points of closest approach ({\tt poca}) to the beamline to convert sparse point clouds of three dimensional data to rich one dimensional data sets amenable to processing by deep neural networks.

The initial algorithm defined a rich, one-dimensional Kernel Density Estimator (KDE) histogram, plus two more one-dimensional histograms, to describe the probabilities of tracks traversing small voxels in space~\cite{Fang:2019wsd,akar2020updated}.
A convolutional Neural Network (CNN) produces a single one-dimensional histogram that nominally predicts Gaussian peaks at the locations of the true PVs using the three input histograms as its feature set. 
A hand-written clustering algorithm then identifies the candidate PVs and their positions. 
The results reported earlier used a ``toy Monte Carlo'' with proto-tracking~\cite{Fang:2019wsd}. 
As discussed below, using track parameters produced by a proposed LHCb Run 3 Vertex Locator (VELO) tracking algorithm~\cite{Hennequin:2019itm} leads to significantly better performance.

The original KDE~\cite{akar2020updated} is a projection of a three-dimensional probability distribution in voxels that has contributions {\em only} when two tracks pass close to each other. 
Calculating this KDE exactly is very time-consuming, and we want to replace it with a deep learning (DL) algorithm of its own. 
Initial attempts suggested that learning track interactions would be difficult, so we defined two new KDEs, one summing probabilities of individual tracks in spatial voxels and one summing squares of probabilities. 
The original KDE used a combination of these. 
As discussed below, using only one of these KDEs as an input feature leads to worse results than using the original KDE, but using both leads to significantly better performance.

Defining KDEs that depend only on individual track parameters, and {\em not} on the direct interaction of two tracks, allows a {\tt tracks-to-KDE} DL algorithm to build a pretty good estimate of the numerically calculated ``probability'' KDE using the ensemble of individual tracks parameters as input features. 
A preliminary model of this DL algorithm was connected to a simple, somewhat limited performance {\tt KDE-to-hists} model to test whether such a {\tt tracks-to-hists} algorithm can achieve as good fidelity as the {\tt KDE-to-hists} algorithm does using a numerically calculated KDEs. Initial results are encouraging.

Modified versions of our original CNN architecture~\cite{Fang:2019wsd} were studied to understand how performance depends on the number of model parameters, the number of network layers, the number of channels per layer, whether batch normalization is used, and the extent to which skip connections are used. 
We refer to these models as members of the {\tt AllCNN} family. 
A one-dimensional version of the U-Net architecture~\cite{ronneberger2015unet}, originally developed for two-dimensional biomedical images, was also tested using toy Monte Carlo and find that it (slightly) outperforms the best of our {\tt AllCNN} models when both are trained with the same KDE.

\section{Performance Evolution}
\label{sec:evolution}

\begin{figure}
\centering
\includegraphics[width=0.55\textwidth,clip]{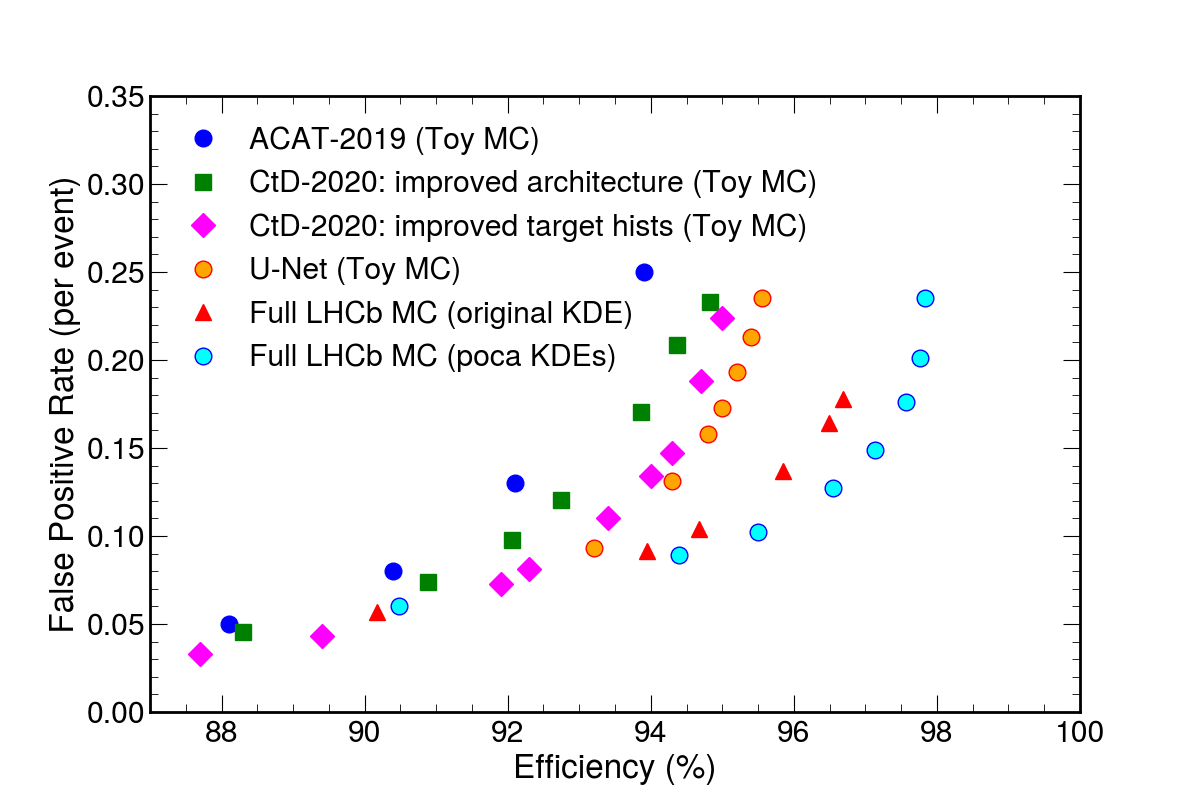}
\caption{
Comparison between the performance of models reported in previous years (labeled ACAT-2019 and CtD-2020) and the new models described in detail in the text. 
An asymmetry parameter between the cost of overestimating contributions to the target histograms and underestimating them is varied to produce the families of points observed.
}
\label{fig:evolution}
\end{figure}

Figure \ref{fig:evolution} shows how the performance of the {\tt KDE-to-hists} algorithms have evolved over time.
The solid blue circles show the performance of any early model described at ACAT-2019~\cite{Fang:2019wsd}.
The green squares and magenta diamonds show the performances described at Connecting-the-Dots in 2020~\cite{akar2020updated}.
The efficiency is shown on the horizontal axis and the false positive rate per event is shown on the vertical axis.
Both quantities are evaluated from a matching procedure done by a heuristic algorithm, based on the PV positions along the beam axis, $z$. 
A predicted PV is matched if the distance, $\Delta z$, between it's position, $z_{\rm pred}$, and the true PV position, $z_{\rm true}$, satisfies $\Delta z = |z_{\rm pred}-z_{\rm true}| \leq 0.5 {\rm mm}$.
The false positive rate is obtained from ratio of remaining predicted PVs after the matching procedure over the total number of true PVs.
An asymmetry parameter between the cost of overestimating contributions to the target histograms and underestimating them~\cite{Fang:2019wsd} is varied to produce the families of points observed.
Those models were trained and tested using the original KDE and toy MC.
The reported results, throughout this document, come from statistically independent validation samples.
The model of the magenta diamonds was tested on full LHCb MC data in which the (original) KDE was derived from a full VELO tracking algorithm~\cite{Hennequin:2019itm}. 
We observed slightly increased performance compared to toy MC (not shown). 
The model was then trained and tested using a full LHCb MC sample yielding significantly better results, shown as the red triangles in Fig.\ref{fig:evolution}.
An algorithm that replaces the original KDE with two KDEs produces even better results, plotted as cyan-filled circles. 
The difference is most pronounced at efficiencies greater than $ \sim 94\% $. The improved performance is primarily associated with correctly finding true lower multiplicity PVs. 
Training the same model as that used to produce the red triangles, but using one of the new KDEs in place of the original KDE, leads to worse performance. 
The algorithm using  both of the new KDEs learns how to combine their information effectively. 
The plot also shows results for a U-Net model~\cite{ronneberger2015unet} using the same toy MC as used to train the ``traditional'' {\tt KDE-to-hists} models. Its performance is better than that of the other models trained and tested on the same data. 
Its architecture is discussed below.

\section{Kernel Density Estimators}
\label{sec:kde}
\noindent
Kernel generation converts sparse three-dimensional data into  a small number of feature-rich one-dimensional data sets.
Deep neural networks (DNNs) can transform these into one-dimensional histograms from which PV candidates are easily extracted.
Each of our one-dimensional data sets consists of 4000 bins along the $z$-direction (beamline), each 100\,$\mu$m wide, spanning the active area of the VELO around the interaction point, such that $z \in [-100,300]\,{\rm mm}$. 
In the original KDE, each $z$ bin of the histogram is filled by the maximum kernel value in $x$ and $y$, where the kernel is defined by
\begin{equation}
\mathcal{K}(x,y,z) = \frac{\sum_\mathrm{tracks}\mathcal{G}(\mathrm{IP}(x,y)|z)^2}{\sum_\mathrm{tracks}\mathcal{G}(\mathrm{IP}(x,y)|z)} -
 \sum_\mathrm{tracks}\mathcal{G}(\mathrm{IP}(x,y)|z)\ .
 \label{eqn:kernel}
\end{equation}
In Eqn.~(\ref{eqn:kernel}), $\mathcal{G}(\mathrm{IP}(x,y)|z)$ is a Gaussian function, centered at $x=y=0$ and evaluated at the impact parameter IP$(x,y)$: the distance of closest approach of a track projection to a hypothesized vertex at position $x,y$ for a given $z$.
The width/covariance of $\mathcal{G}$ is given by the IP$(x,y)$ uncertainty/covariance matrix.
Finding the maximum $\mathcal{K}(x,y,z)$ is a two step process. 
Kernel values are first computed in a coarse $8\times 8$ grid in $x,y$; the parameters of that search are then taken as starting points for a \texttt{MINUIT} minimization process to find the maximum kernel value for each bin of $ z $. 
The values of $ x $  and $ y $ where the kernel is maximum are saved as secondary feature sets {\tt xMax} and {\tt yMax}

Equation~(\ref{eqn:kernel}) is computed from the location, direction, and covariance matrix of input tracks.
This information can be provided by a standalone toy Monte Carlo proto-tracking~\cite{Fang:2019wsd} or from a proposed
LHCb Run\,3 production Velo tracking~\cite{Hennequin:2019itm}.
While the former uses a heuristic Gaussian width to estimate the IP$(x,y)$ uncertainty, the latter  uses the measured covariance matrix of the track state closest to beamline to compute the IP$(x,y)$ covariance.
To reduce the computational load, the KDE contributions are only calculated in regions near points where two tracks pass within $ 100 \, \mu$m of each other.
This is completely safe in terms of finding true PV positions, but leads to discontinuities in KDE values as a function of $ z $.

\begin{figure}
\centering
\includegraphics[width=0.75\textwidth,clip]{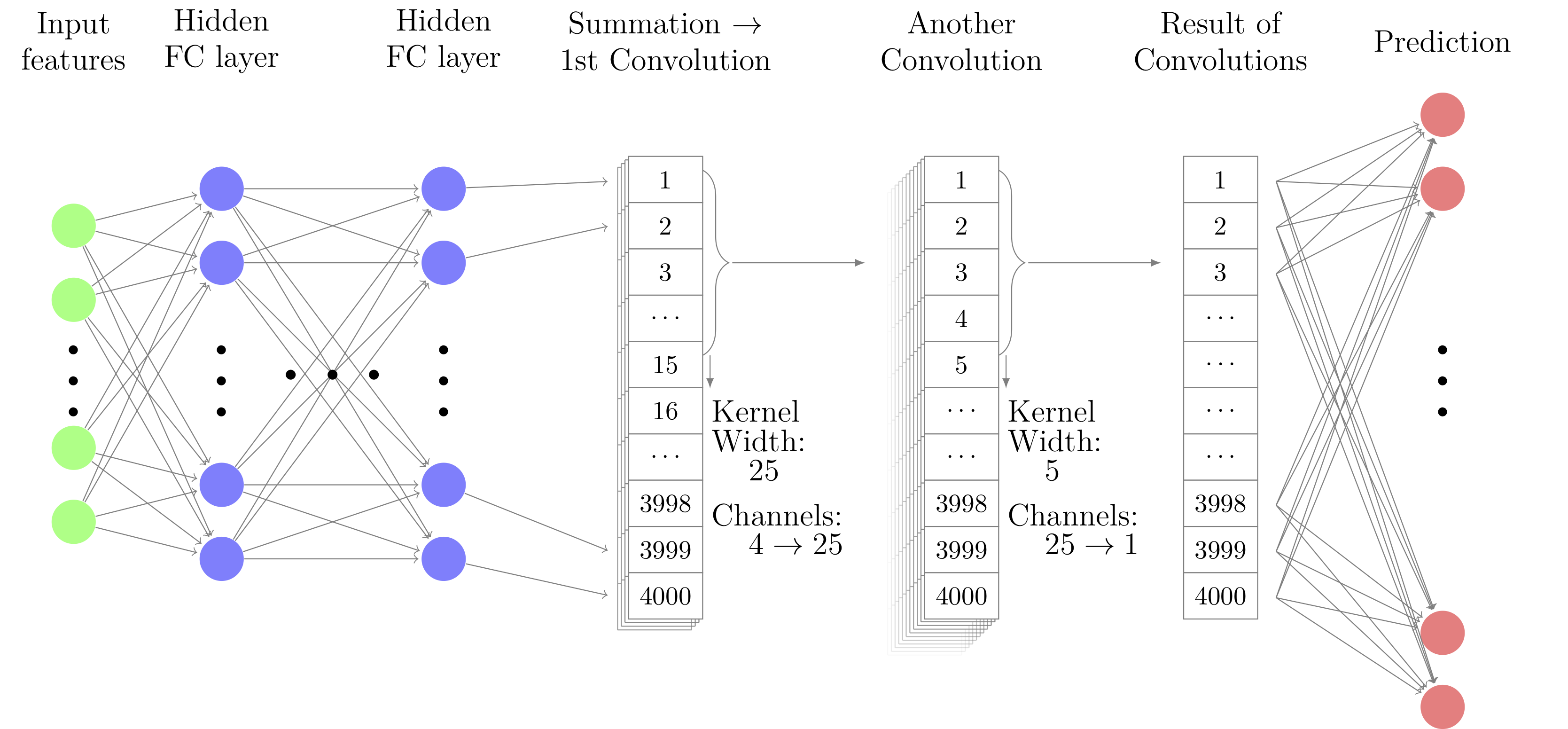}
\caption{
This diagram illustrates the deep neural network used to predict an event's {\tt KDE-A} from its tracks' {\tt poca-ellipsoid}s.
Twelve fully connected layers (all but the last with 50 neurons) populate four 4000-bin channels in the last of these layers, for each track. 
These contributions are summed and processed by two convolutional layers that provide a mechanism for the tracks to interact. 
A final fully connected layer was required for the learning to converge.
}
\label{fig:DDplus}       
\end{figure}

A first attempt to design an algorithm to predict KDE distributions from track parameters led to relatively poor performances.
It was designed to treat each track's contribution to the KDE separately. 
Adding the explicit requirement that a track contribute to the KDE only when near another track allowed to increase the algorithm performances. 
More specifically, the original KDE was replaced by a new one, {\tt KDE-A}, based on each track's position of closest approach ({\tt poca}) to the beamline and the error ellipsoid defined by
\begin{equation}
   A (\Delta x)^2 + B  (\Delta y)^2 + C (\Delta z)^2 + 2 
     (D \Delta x \Delta y + E \Delta x \Delta z + F \Delta y \Delta z) = 1
\end{equation}
where $ \Delta x = x - x_{\rm poca} $, etc.
At any point in space, a Gaussian probability associated with this track's {\tt poca}-ellipsoid is calculated as
\begin{equation}
  {\cal G}( x, y, z)  = \frac{{\rm exp} \left ( -0.5  (
    A (\Delta x)^2 + B  (\Delta y)^2 + C (\Delta z)^2 + 2
     (D \Delta x \Delta y + E \Delta x \Delta z + F \Delta y \Delta z) \right )}
   {(\mbox{ellipsoid volume}){}^{3/2}}
\end{equation}
For each voxel, the {\tt KDE-A} contribution is calculated as the sum of {\tt poca}-ellipsoid probabilities:
\begin{equation}
{\cal K}_A = \sum_\mathrm{tracks} {\cal G}( x_i, y_i, z_i) \, .
\end{equation}
where $ i $ denotes  the track index.
For each bin of $ z $, the largest value of $ {\cal K}_A $ observed in an $ (x, y, z) $ voxel is projected out as the {\tt KDE-A} value at that value of  $ z $.
However, the performance of the {\tt KDE-to-hists} algorithm using {\tt KDE-A} in place of the original KDE was significantly worse.
As the definition of the original kernel in Eqn.~\ref{eqn:kernel} combined linear and quadratic terms in  $ \mathcal{G}(\mathrm{IP}(x,y)|z) $, we defined {\tt KDE-B} using a procedure parallel to that for defining {\tt KDE-A} but replacing the linear sum of probabilities in $ {\cal K}_A $ with 
\begin{equation}
\label{eqn:KA}
{\cal K}_B = \sum_\mathrm{tracks} {\cal G}( x_i, y_i, z_i)^2 \, .
\end{equation}
Using both {\tt KDE-A} and {\tt KDE-B} as input features, in place of the original KDE, improved the performance of the {\tt KDE-to-hists} algorithm, as noted in the earlier discussion of Fig.~\ref{fig:evolution} (see the red triangles and cyan-filled circles).

\begin{figure}[t!]
\includegraphics[width=0.50\textwidth,clip]{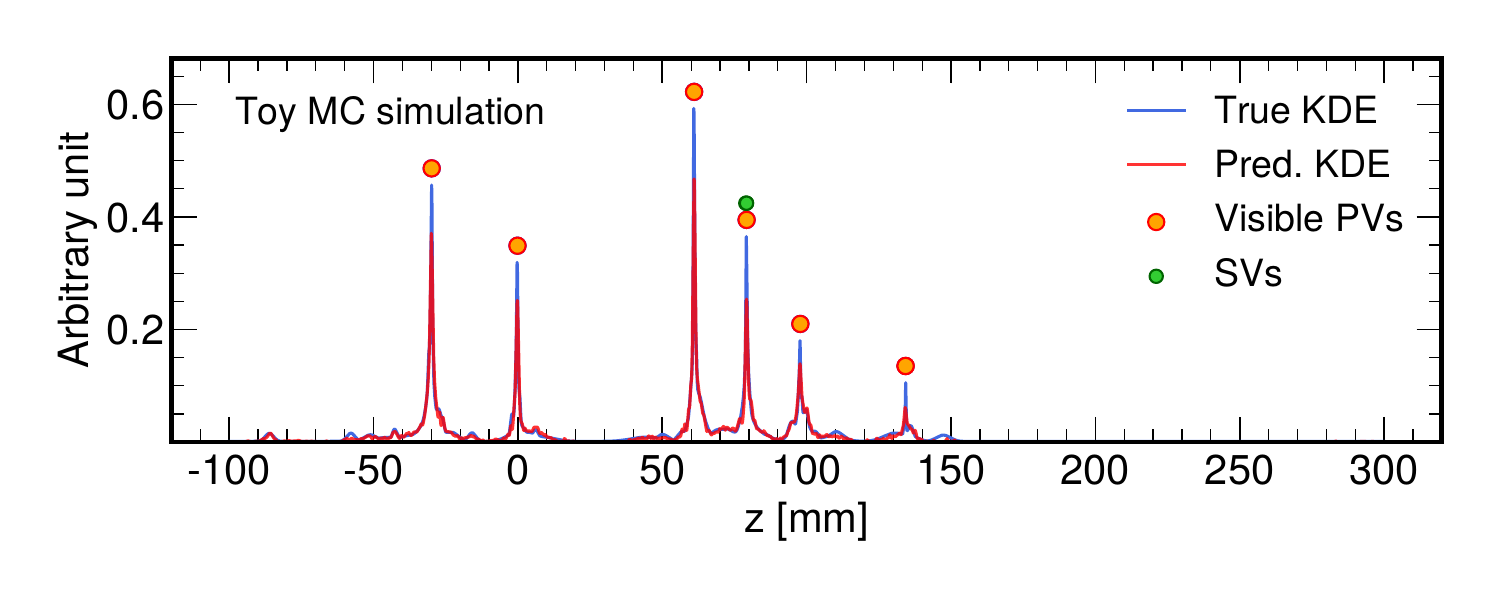}
\includegraphics[width=0.50\textwidth,clip]{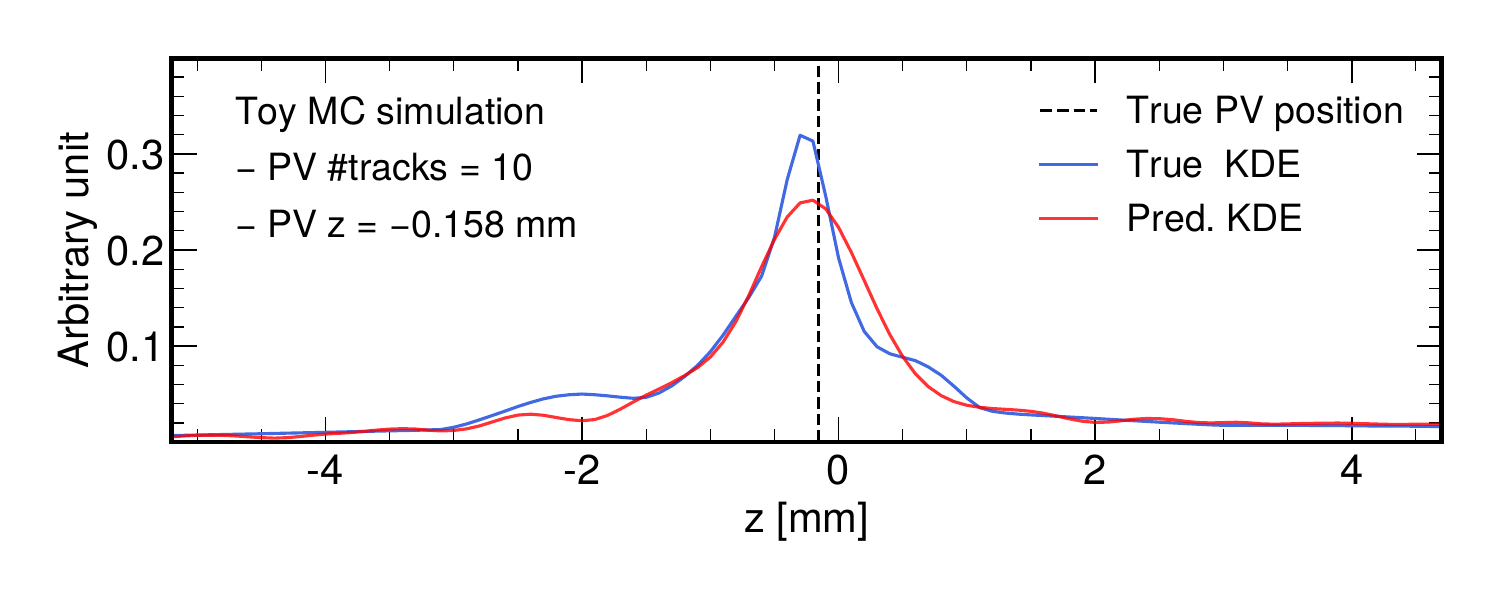}
\caption{
The plot on the left illustrates  {\tt KDE-A} (shown in blue), calculated as described in the text following Eqn.~\ref{eqn:KA}, and the KDE predicted by the DNN model described in the subsequent paragraph (shown in red) for a single event. 
Each of the 4000 bins in the histogram corresponds to a range in the $ z $-direction of $ 100 \,  \mu$m.
The plot on the right shows a zoomed-in view of the 50 bins on either side of the  location of the second PV in that event, whose location is denoted by the dashed vertical line.
} 
\label{fig:iter21Event}       
\end{figure}

While solely predicting {\tt KDE-A} distributions from tracks parameters with reasonably good fidelity using a DNN is quite strait-forward, achieving satisfactory performance {\em does} require a mechanism that encourages the predictions for each track to interact with the predictions for the other tracks.
This was studied using toy MC rather than full LHCb MC as the former data sets are larger than the latter.
For these, the model architecture was not optimized -- the choices of the numbers of layers of each type, the number of channels per layer, the kernel sizes in the convolutional layers, etc., were chosen almost arbitrarily.
With these caveats, the architecture is illustrated in Fig.~\ref{fig:DDplus}.
Originally, there were 12 fully connected layers that predicted {\tt KDE-A} contributions for each track separately and then summed them.
The performance of this model was mediocre (at best). 
To allow the tracks to interact, the single 4000-bin channel in the 12th layer that was originally meant to be the KDE prediction was replaced by four latent 4000-bin channels. 
These are used as input to two convolutional layers (and one more fully connected layer) that produce a final 4000-bin prediction. 
This more complicated model was trained starting with the weights and biases of the first 11 fully connected layers fixed while those of the new layers were learned.
Once the full model began to display a semblance of performance, all the model parameters were floated and learned over many epochs, using progressively larger training samples. 
Fig.~\ref{fig:iter21Event} compares {\tt KDE-A} (the blue histogram) for the first event in the validation sample with the corresponding KDE predicted by the model (the red histogram) after about 300 hours of training on an nVidia RTX2080Ti using {\textpack PyTorch}.
Qualitatively, the predicted KDE reproduces the macroscopic features of {\tt KDE-A}, as seen in the plots on the left.
On a finer scale, there are differences, as seen in the plot on the right.
This shows 50 bins on either side of the location of the first PV in the event.
The predicted KDE peaks at the location of the PV, as does {\tt KDE-A}, but its height is lower and its width is greater.
This is typical of the learned KDE, although there is tremendous variation from the region about one  true PV in an event to another and from event to event.
A mean-squared-error function comparing the predicted KDE histogram to the  {\tt KDE-A} is used to train the model.
But the real goal is using a well-trained {\tt tracks-to-KDE}  model as a step towards defining a {\tt tracks-to-hist} model that performs at least as well as the best model in Fig.~\ref{fig:evolution}.

As a proof-of-principle, we first trained a {\tt KDE-to-hists} model using {\tt KDE-A} as the only input feature set.
The efficiency and false positive rate were \textcolor{black}{94.3\% and 0.27 per event}.
We then merged the {\tt tracks-to-KDE} model described above with this {\tt KDE-to-hists} model to create a {\tt tracks-to-hists} model.
With the weights of the combined model set to be those of the independent models, the efficiency and false positive rate were approximately \textcolor{black}{88\% and 0.88 per event}.
We then allowed the combined model to learn improved weights for the {\tt tracks-to-KDE} layers while the weights for the {\tt KDE-to-hists} layers were held constant and vice versa.
In these cases, the improvements were very slow, with the efficiencies increasing a (small) fraction of a percent and the false positive rates improving by less than 0.02 over the course of several days training on nVidia RTX 2080Ti GPUs.
Allowing all weights and biases in the combined model to float produced significantly improved performance. 
After several days of training, the efficiency increased to 91.0\% and the false positive rate dropped to 0.80 per event before improvement completely plateaued.

As noted above, the predicted KDEs tend to produce broader features in the vicinity of true PVs, as seen in Fig.~\ref{fig:iter21Event}.
When increasing the threshold used in the matching procedure from $ 500 $ to $ 750 \, \mu$m, we observed a slight rise in efficiency, to 93.7\%, and a drop in false positive rate to 0.69 per event.
Although these efficiencies (original and alternative definitions) are not as high as that of the {\tt KDE-to-hists} model, and the false positive rates are higher, the fact that simultaneously learning the combined model parameters produces significantly better results than simply connecting the two networks with their independently learned parameters suggests that it should be possible to create a highly performant {\tt tracks-to-hists}  model. 
It also suggests that improving the {\tt tracks-to-KDE} model so that it produces more narrow peaks around PV positions will be necessary to significantly improve the merged model performance.

\section{Alternative Model Architectures}

\begin{figure}
\centering
\begin{minipage}[t]{0.48\linewidth}
\includegraphics[width=1.00\textwidth,clip]{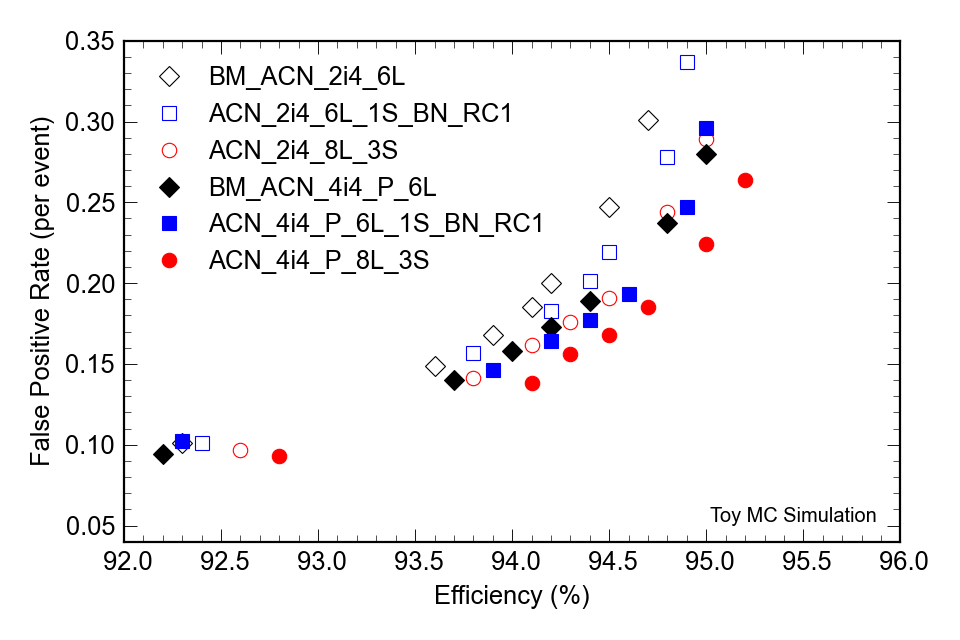}
\caption{
This plot illustrates the improved performance of several models when {\tt xMax} and {\tt yMax} are added as perturbative features, as discussed in the text.
For each  model, the open markers show the efficiency versus false positive rate as the cost function asymmetry parameter is varied, while filled markers show the performance when the perturbative features are added.
Details on the model variations are given in the appendix.
} 
\label{fig:compare-perturbative}       
\end{minipage}
\quad
\begin{minipage}[t]{0.48\linewidth}
\includegraphics[width=1.00\textwidth,clip]{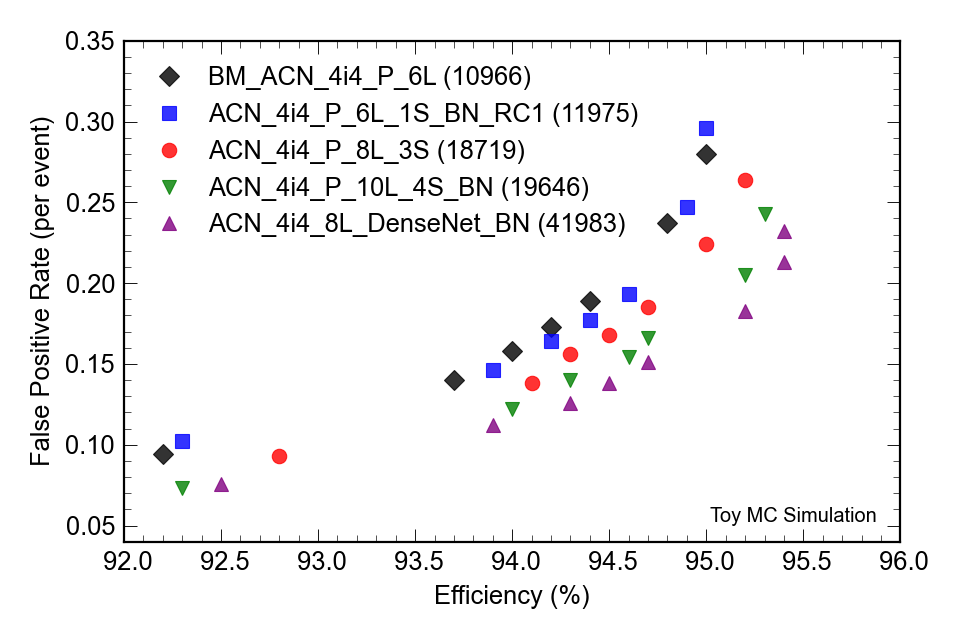}
\caption{
This plot illustrates the  performance of several models built from a benchmark model with variations described in the text.
The labels for each model encode the nature of the variations, as described in Table~\ref{tab:tags} in the appendix.
The number of parameters learned by each model is shown in parentheses to the right of its  label.
\vskip 0.42in
} 
\label{fig:pacnn}       
\end{minipage}
\end{figure}

\noindent
The architectures of the models whose performance is discussed in Sec.~\ref{sec:evolution} and illustrated in Fig.~\ref{fig:evolution} are similar, but vary as discussed at conferences last year~\cite{akar2020updated} and above.
We have now investigated some of these variations, and others, more systematically.
We have studied a qualitatively new architecture as well.
We find that adding the secondary feature sets {\tt xMax} and {\tt yMax} to the models ``perturbatively'' systematically improves performance.
We also find that models with more parameters generally out-perform similar models with fewer parameters. 
From a computer science perspective, the best performance may be considered optimal.
However, when deploying software for use in a HEP experiment, use of memory and processor time may be considered as well.
For example, it may be appropriate to trade off small gains in performance if doing so requires only half the computing resources.

The results of adding perturbative features to three baseline models is illustrated in Fig.~\ref{fig:compare-perturbative}.
Once a model has been fully trained using only the KDE feature set, its weights and biases are fixed and a second network is trained using only the secondary features; the final prediction is the product of the two networks' outputs (or generated from a sum or concatenation of their outputs).
After the secondary network is trained in this way, all the weights and biases of the combined model are trained simultaneously to produce the final model.  
We find that initially separating the baseline training from the secondary feature training is necessary for deep learning to start.
In each case shown (and others not shown to avoid visual congestion), adding {\tt xMax} and {\tt yMax} features to the model perturbatively increases the efficiency {\em and} reduces the false positive rate.

The benchmark model used as the starting point for studies of alternative models has six convolutional layers, very similar to the model with four convolutional layers illustrated in Fig.~\ref{fig:DDplus} of Ref.~\cite{Fang:2019wsd}.
In the results presented below, {\tt xMax} and {\tt yMax} have been added as perturbative features.
The primary goal of these studies was to understand how the fidelity of a model varies with the number of parameters and the use of  skip connections, batch normalization, etc.
Table~\ref{tab:tags} in the appendix lists the types of variations that were considered.
Results from 5 of these models are shown in Fig.~\ref{fig:pacnn}.
In general, the larger the number of parameters, the greater the fidelity of the model.
That with the best performance is an adaptation of the \textbf{DenseNet} architecture~\cite{huang2018densely} which connects each layer to every other layer in a feed-forward fashion, the ultimate limit adding skip connections.
Note that these  studies were done in parallel with others reported here.
We have not yet incorporated what was learned here into, for example, the best performing models of Fig.~\ref{fig:evolution}.

\begin{figure}[t]
\centering
\includegraphics[width=0.60\textwidth,clip]{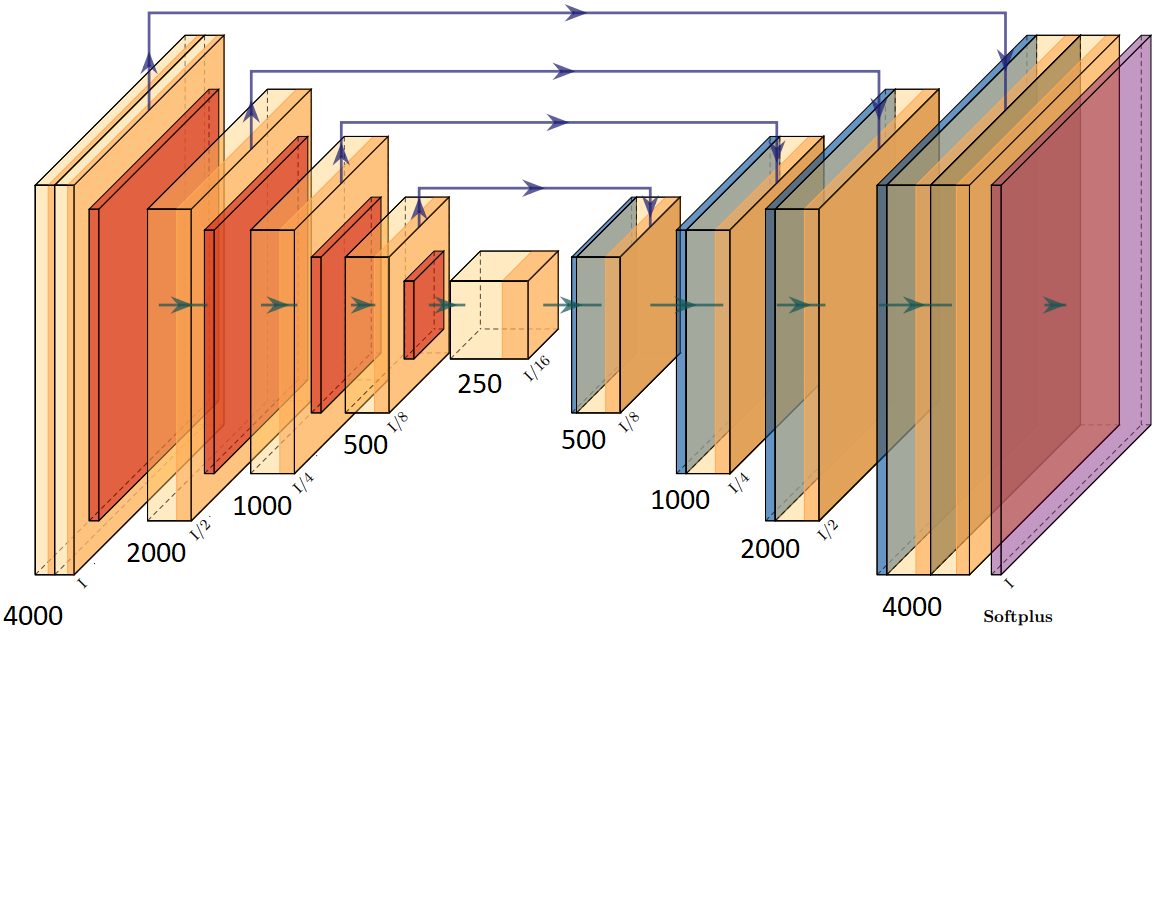}
\vspace{-1.5cm}
\caption{
This is the architecture of a U-Net model modified for use in our one-dimensional {\tt KDE-to-hists} problem.
The orange blocks are convolutional layers followed by ReLU activation functions, the red blocks are down-sampling operations, and the blue blocks are up-sampling operations.
The four arrows that join  down-sampling layers to  up-sampling layers of the same dimensionality denote skip connections.
} 
\label{fig:unet}       
\end{figure}

Inspired by its success solving other tasks that require shape-preservation between input and output (object detection, semantic segmentation), we have modified the popular U-Net architecture~\cite{ronneberger2015unet} for our one-dimensional problem. 
A diagram of the architecture is shown in Fig.~\ref{fig:unet} -- orange blocks are convolutional layers followed by ReLU activation functions, red blocks are down-sampling operations (max pooling with kernel size and stride~2), and blue blocks are up-sampling operations (transposed convolutions with kernel size and stride~2). 
The arrows represent the flow of information throughout the network.
It differs from a traditional autoencoder as some information bypasses successive layers and is added back in later stages via a concatenation or addition operation (a skip connection). 
We investigated this architecture for a number of reasons.
Conceptually, it captures a receptive field that scales as $2^n$, where $ n $ is number of pooling operations. 
This can be compared to the {\tt AllCNN} architectures where the receptive field scales linearly with $ n $.
We hypothesize that the added ease in down-sampling is important to achieve the granular resolution we want. 
In addition, we hypothesize that the addition of skip connections up to the second-to-last layer in the network allows the network to ``remember'' fine-grained detail about the input, and more effectively localize primary vertices. 
Experiments show that removing skip connections one-by-one consistently degrades results by a minor amount. 
Including all four skip connections, the architecture achieves an efficiency of 95\% with a false-positive rate of 0.16 per event, slightly better than the best {\tt AllCNN} architecture using the same toy MC data and features.
We anticipate that this architecture will perform at least as well as the best {\tt AllCNN} architecture when trained using full LHCb MC plus  {\tt KDE-A} and {\tt KDE-B}  as input features.

\section{Summary and Conclusions}

Since we  presented results at Connecting-the-Dots 2020, we have tested  {\tt AllCNN} {\tt KDE-to-hists} models using full LHCb Monte Carlo rather than toy Monte Carlo and find they are (i) more performant without re-training and (ii) even more performant with re-training.
If we replace the original KDE with a pair of KDEs, each built from individual track contributions with no pairwise interactions used explicitly, the performance is better again.
We have built a deep learning {\tt tracks-to-KDE} algorithm that reproduces the new KDE of Eqn.~\ref{eqn:KA} with sufficient fidelity that a merged {\tt tracks-to-hist} algorithm achieves almost as high efficiency as the underlying {\tt KDE-to-hist} algorithm, albeit with worse resolution and a significantly higher false positive rate. 
We conclude that the overall approach is sound, but we need to construct and train a better {\tt tracks-to-KDE} algorithm.

We have also performed systematic studies of alternative {\tt AllCNN} models and a very different model inspired by U-Net.
The results of these studies indicate that the capacity of a model to learn increases with the number of parameters and they suggest specific approaches for improving the performance of the best {\tt KDE-to-hist} model so far.
The very similar performance of the model inspired by U-Net and the best {\tt AllCNN} model trained on the same KDE invites the question of how two such different architectures have ``learned'' the same ``concepts'' and whether such ``understanding'' can be generalized.

Two physics insights/hypotheses underly our current approach to developing hybrid deep learning algorithms for identifying and locating primary vertices.
First, tracks from primary vertices pass close enough to the beamline that each can be characterized by its {\tt poca-ellipsoid}.
Second, the sparse three-dimensional information encoded in these ellipsoids can be transformed to  rich, one-dimensional KDEs that can, in turn, serve as input feature sets for relatively simple deep neural networks.
These can be trained on large samples of simulated events (hundreds of thousands) to produce one-dimensional histograms that can be processed by traditional, heuristic algorithms to extract information for the next stage of event reconstruction and classification.

Our specific algorithms are being designed for use in LHCb with its Run 3 detector.
As the salient characteristics exist in the ATLAS and CMS experiments, as well, appropriately modified versions might work for them.
A heuristic ATLAS vertexing algorithm~\cite{ATLAS:2019jmx, Sanderswood:2019rta}, already uses some conceptually similar ideas, as does an algorithm~\cite{Bocci:2020pmi} designed for the CMS detector, upgraded for operation in the high luminosity LHC era.
Developing and deploying machine learning inference engines that are highly performant and satisfy computing system constraints requires sustained effort.
The results reported here should encourage work focussed on using deep neural networks for identifying vertices in high energy physics experiments.

\section{Acknowledgments}
\noindent

The authors  thank the LHCb computing and simulation teams for their support and for producing the simulated LHCb samples used in this paper.
The authors  also thank the full LHCb Real Time Analysis team, especially the developers of the VELO tracking algorithm~\cite{Hennequin:2019itm} used to generate the ``full LHCb MC'' results presented in Fig~\ref{fig:evolution}.

This work was supported, by the U.S. National Science Foundation under Cooperative Agreement 
OAC-1836650 and awards 
PHY-1806260, 
OAC-1739772, and 
OAC-1740102.

\newpage

%

\bibliography{vchep.bib}
\vskip 0.2in
%
\clearpage
\appendix
\section*{Appendix: Models and Tags}

\begin{figure}[htb]
\centering
 \includegraphics[width=0.70\textwidth,clip]{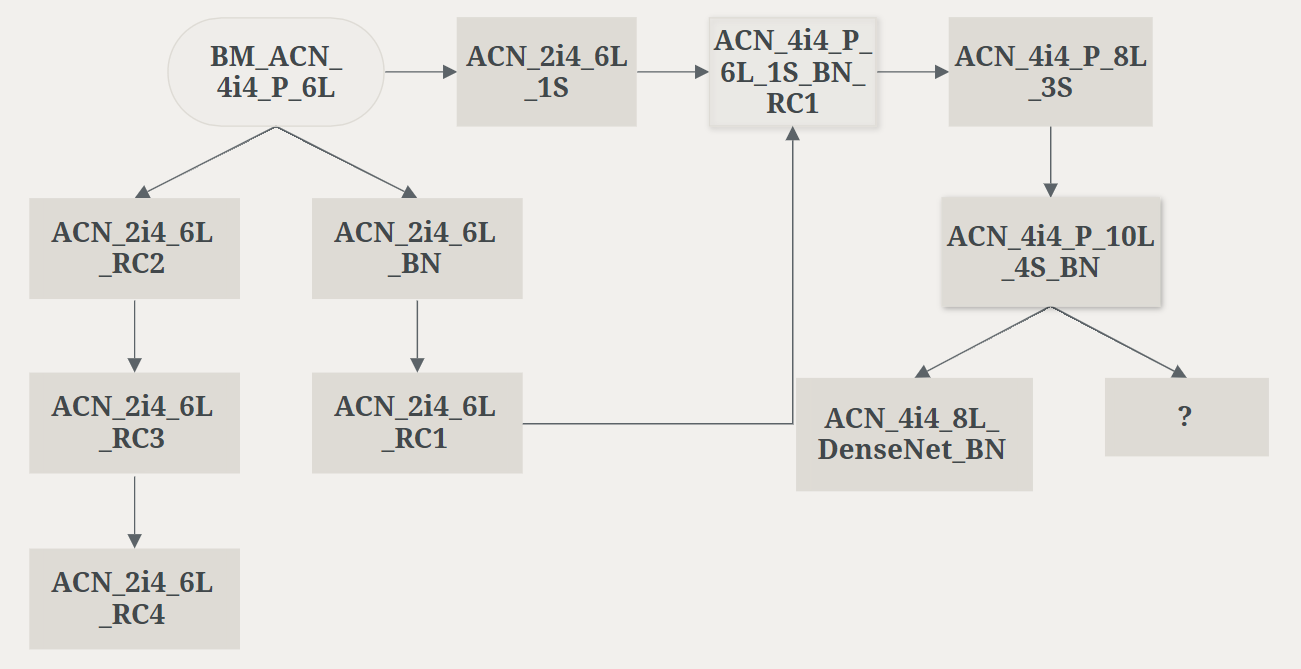}
 \caption{
  This figure shows how models studied relate to each other.
  The tags indicating the variations in the architectures
  are described in Table~\ref{tab:tags}.
  } 
 \label{fig:models} 
\end{figure}

The models discussed in the studies of systematic variations are derived from a benchmark {\tt AllCNN} model.
The relationships between these models are shown in Fig.~\ref{fig:models}.
As indicated in Table~\ref{tab:tags}, the numbers of layers, channels per layer, use of skip connections, batch normalization, etc., could be varied.
Not all combinations were studied, but the conclusion presented in the body of the text -- that increasing the numbers of parameters  increases the capacity of a network to learn well, almost independently of their origins -- seems to be robust.

These are the tags used to label each model name. The tags determine some of the key features of the model at a high level.
\begin{table}[ht]
\begin{center}
\caption{Neural network model tags for names}
\label{tab:tags} 
\begin{tabular}{cl} 
\hline
\multicolumn{1}{c}{Tag} & \multicolumn{1}{l}{V (Meaning)} \\
\hline
\#S &   number of skip connections, if any \\
\#L &   number of layers \\
BN &   Batch Normalization \\
ACN &  AllCNN ``family'' of models \\
\#i\# &  step number out of total steps \\
RC\# &  reduced channel size, followed by iteration number  \\
IC\# &  increased channel size, followed by iteration number \\
RK\# &  reduced kernel size, followed by iteration number \\
IK\# &  increased kernel size, followed by iteration number \\
C &    concatenation of perturbative and non-perturbative layers at end \\
BM &   benchmark; changes made to future models based on tagged reference model \\
\hline
\end{tabular}
\end{center}
\end{table}

The tag hierarchy is of the format:

\smallskip
{BM-ACN-\#i\#-P-\#L-\#S-BN-RC-\#-IC\#-RK\#-IK\#-C}

\end{document}